\documentclass[vecphys]{svmult}

\usepackage{makeidx}         
\usepackage{graphicx}        
\usepackage{multicol}        
\usepackage[bottom]{footmisc}
\usepackage{natbib} 
\makeindex             
\newcommand\aap{{A\&A}}%
\newcommand\aaps{{A\&AS}}%
\newcommand\apj{{ApJ}}%
\newcommand\mnras{{MNRAS}}%
\newcommand\aj{{AJ}}%
\newcommand\apjl{{ApJ}}%

\begin{document}

\title*{The Warp and Spiral Arms of the Milky Way}
\author{E.S. Levine\inst{1}\and Leo Blitz\inst{1} \and Carl Heiles\inst{1}\and
Martin Weinberg\inst{2}}
\institute{UC Berkeley Astronomy Dept, 601 Campbell Hall, Berkeley CA 94720 USA
\texttt{elevine@astron.berkeley.edu}
\and University of Massachusetts Amherst Astronomy Dept, LGRT-B 619E, 710 North Pleasant Street, Amherst MA 01003 USA}
%
%
\maketitle

We examine the outer Galactic HI disk for deviations from the $b=0^\circ$ plane by constructing maps of disk surface density, mean height, and thickness. We find that the Galactic warp is well described by a vertical offset plus two Fourier modes of frequency 1 and 2, all of which grow with Galactocentric radius. A perturbation theory calculation demonstrates that the tidal influence of the Magellanic Clouds creates distortions in the dark matter halo, which combine to produce disk warp amplitudes comparable to the observations. We also construct  a detailed map of the perturbed surface density of HI in the outer disk demonstrating that the Galaxy is a non-axisymmetric multi-armed spiral. Overdensities in the surface density are coincident with regions of reduced gas thickness.

\section{Method}\label{method}
We use the 21cm Leiden/Argentine/Bonn (LAB) data \citep{LAB,HB1997,BALMPK2005,ABLMP2000} to conduct a quantitative study of the warp and spiral structure. We used the Hanning smoothed data, which have a velocity resolution of 1.9 km s$^{-1}$, and restricted our sample to $|b|\le 30^\circ$. We apply a median filter to the data to remove angularly small, bright features that are clearly not associated with the disk.  To create a map of the Galaxy from the data, we must assume a rotation structure for the disk. Using a flat rotation curve with circular orbits results in a large asymmetry between the surface densities at Galactic longitudes on either side of $\ell=0^\circ$ and $\ell=180^\circ$. To eliminate this discontinuity, we assume that the gas is traveling on elliptical streamlines configured such that the surface densities are smooth across the $\ell=180^\circ$ line (see \citet{LBH2006a}). We exclude all points that lie within $345^\circ\le\ell\le15^\circ$ or $165^\circ\le\ell\le195^\circ$. Points in these two wedges have velocities along the line of sight that are too small with respect to their random velocities to establish reliable distances. 

With this rotation structure, we transform the data to a three dimensional density grid, $\rho(R,\phi,z)$, where $R$ is Galactocentric radius and $z$ is height off the plane. The Galactocentric azimuth $\phi$ is defined such that it converges with Galactic longitude $\ell$ at large $R$. We use a dispersion filter on $\rho$ to remove extended clouds near the disk or spurs that have split off from the disk. Finally, we calculate the surface density $\Sigma$ and mean height $h$ as a function of position $(R,\phi)$ in the disk. The half-thickness $T_h$ is calculated from $\rho$ before the dispersion filter is applied, following the procedure described in \citet{HJK1982}.

\section{Warping}
\subsection{Modes}
\begin{figure}
\centering
\includegraphics[height=6.3cm]{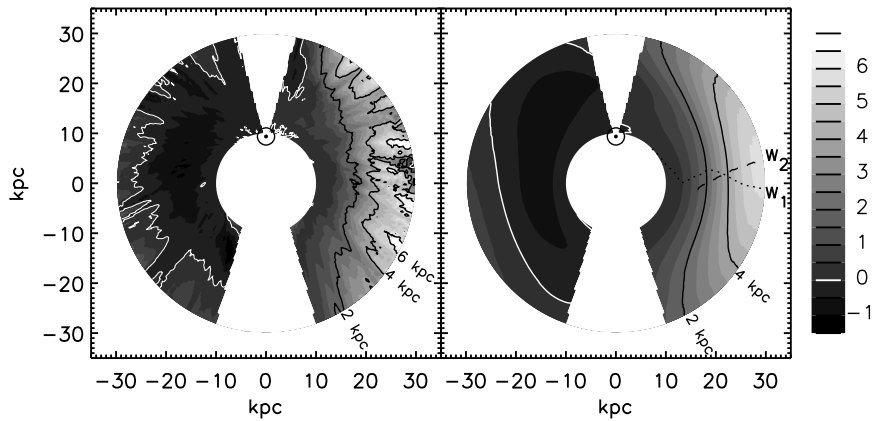}
\caption{\label{fig:warp}Left panel: $h(R,\phi)$. Right panel: The fit to the warp is plotted, along with the lines of maximum amplitude for the m=1 (dotted) and m=2 (dashed) modes. These lines are marked $W_1$ and $W_2$ respectively. The white contour line denotes a height of 0 kpc; black lines mark the 2, 4, and 6 kpc elevations. The colorbar is marked in kpc.}
\end{figure}

A Lomb periodogram analysis of each radius ring reveals that the power in each of the frequency $m=0$, 1, and 2 modes is consistently larger than that in any other mode for $R> 20$ kpc. Accordingly, we characterize the warp by an offset in the $z$ direction, plus two Fourier modes with frequency $m=1$ and 2. We fit each ring with the function
\begin{equation}
W(\phi)=W_0+W_1\sin(\phi-\phi_1) +W_2 \sin(2\phi-\phi_2);
\end{equation}
this is similar to a fit described in \citet{BM1998}. Each of the three amplitudes $W_i$ and two phases $\phi_i$ in this fit is a function of radius, because we fit each radius ring independently. This three component fit does a good job of reproducing the large-scale features in the mean height map (Fig.~\ref{fig:warp}).

The three amplitude parameters each increase monotonically, with the $m=0$ mode possibly reaching an asymptotic value near the far end of our radius range. At $R\approx11$ kpc, the $m=1$ mode dominates the shape of the warp; the other two modes do not become important until $R\approx 15$ kpc. There is little evidence for precession in the lines of maxima for the $m=1$ and 2 modes, and the line of maxima for the $m=1$ mode is roughly aligned with one of the lines of maxima of the $m=2$ mode.

\subsection{Mechanism}
Many efforts have been directed toward understanding the warp on a theoretical basis. Bending modes have long been suspected as the mechanism creating and maintaining the warp \citep[and many others]{HT1969,SC1988,S1995}, but several other mechanisms have been suggested \citep{KG2000}. Gravitational interaction with satellites such as the Magellanic Clouds has been a long-standing possibility \citep{B1957,W1998}, but there is a debate over whether tidal effects are strong enough to produce the observed effect \citep{K1957}.

We perform a linear perturbation analysis to model the disk's response to the Clouds \citep{WB2006}. We allow the disk to feel the tidal field from the Clouds directly as well as the force from the dark-matter halo wake excited by the Clouds. We adopted the LMC mass from \citet{W1997}, the radial velocity and proper motion from \citet{KVAACDG2006}, and the distance modulus from \citet{F2001}. The warp is a very dynamic structure based on the temporal evolution of the model. This can be seen in the AVI file of the simulations which can be found at \textbf{http://www.astro.umass.edu/$\sim$weinberg/lmc}. Rather than a static structure that might be expected for a warp in response to a triaxial halo, a warp that results from the Clouds is continuously changing shape because of the varying amplitudes and phases of the various modes. As in the data, the $m=1$ mode in the calculations is the strongest and increases nearly linearly out to the edge of the disk. The calculations also show a weak response of $m=0,2$ modes out to about 15 kpc, and then increasing nearly linearly, but with approximately the same amplitude for both.  All of the major features of the warp are reproduced by the calculations, though there are some relatively minor differences which are likely to be due to uncertainties in the modeling. 

\section{Spiral Structure}
Mapping the Milky Way's spiral structure is traditionally difficult because the Sun is imbedded in the Galactic disk; absorption by dust renders optical methods ineffective at distances larger than a few kpc. Radio lines like the 21 cm transition are not affected by this absorption, and are therefore well-suited to looking through the disk. In \S \ref{method}, we described the calculation of $\Sigma(R,\phi)$; we now perform a modified unsharp masking procedure on this map \citep{LBH2006b}. By subtracting a blurred copy from the original image, this technique emphasizes low contrast features in both bright and dim regions. In previous maps \citep{HJK1982} spiral arms were difficult to pick out on top of the global falloff of surface density with radius.

\begin{figure}
\centering
\includegraphics[height=12cm,angle=90]{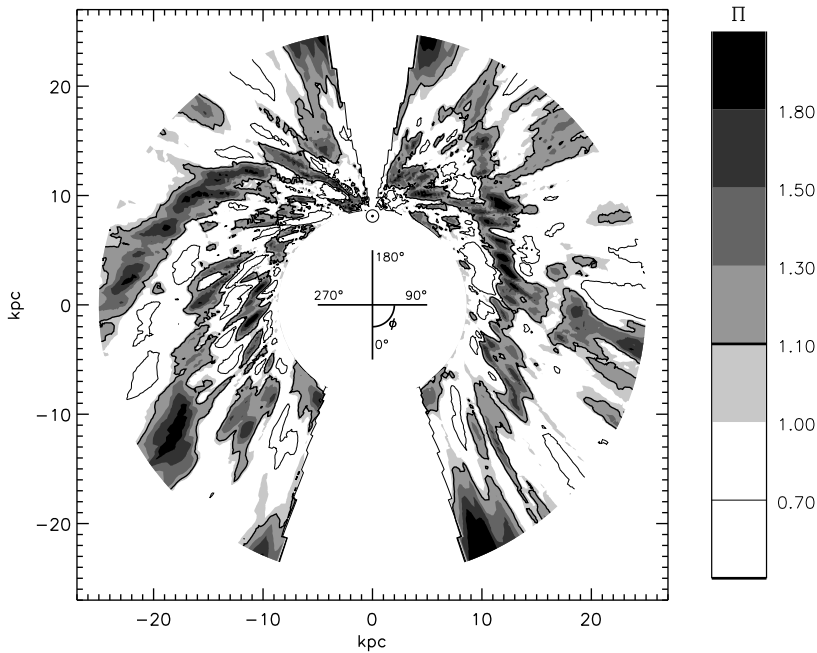}
\caption{\label{fig:spiral}A contour plot of the ratio of the surface density to the local median surface density $\Pi(R,\phi)$. Shaded regions are overdense compared to the local median. The thick solid contour marks the line $\Pi=1.1$, while the thinner contour marks the line $\Pi=0.7$. The values of $\Pi$ for the different contour levels are given by the colorbar.}
\end{figure}

We construct a ratio $\Pi(R,\phi)$ between the surface density and the local median surface density. This is a dimensionless quantity, and therefore is a direct measure of the strength of the surface density perturbations (Fig.~\ref{fig:spiral}). The vast majority of points have values in the range 0.6 to 1.8, implying that the arm-to-interarm surface density ratio is about 3. Several long spiral arms appear clearly on the map, but the overall structure does not have the origin reflection symmetry of a ''grand-design'' spiral. Easily identified spiral arms cover a larger area in the south than in the north; there are three arms in the southern half of the diagram. Each of these arms has already been detected in previous maps out to 17 kpc for the two arms closer to the anticenter \citep{HJK1982} and out to 24 kpc for the `outer' arm \citep{MDGG2004}.

\begin{figure}
\centering
\includegraphics[height=12cm,angle=90]{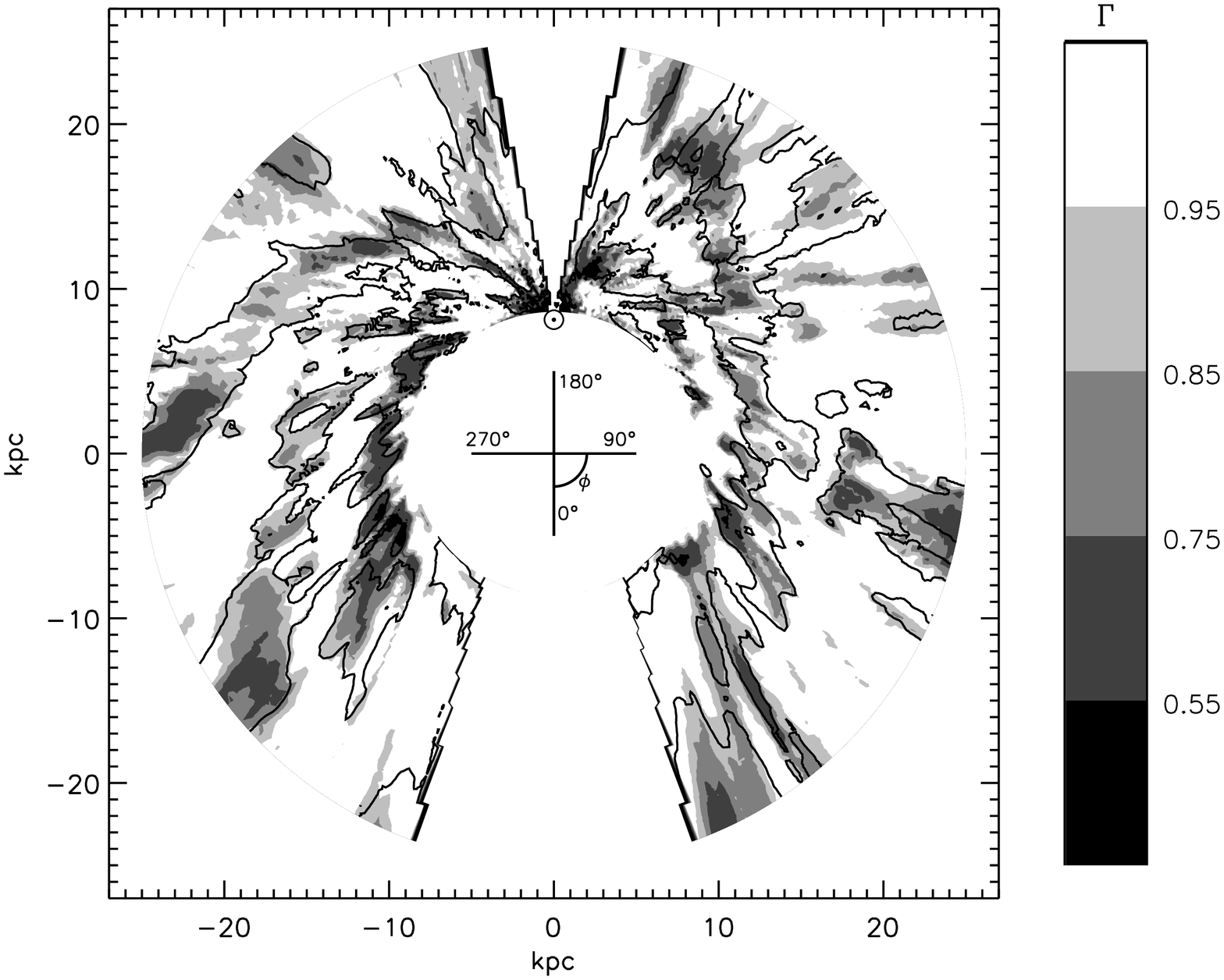}
\caption{\label{fig:thick}A contour plot of the perturbations in the gas thickness, $\Gamma(R,\phi)$. Shaded regions have reduced thickness compared to the local median. The perturbation levels for the different contour levels are given by the colorbar. The solid contour marks the line $\Pi=1.1$, the same as in Fig.~\ref{fig:spiral}, to show the alignment of the overdense surface densities with the thinner gas regions.}
\end{figure}

We also perform the modified unsharp masking on the thickness $T_h(R,\phi)$; spiral structure is evident in this map as well. We plot the $\Pi=1.1$ overdensity contour from the surface density perturbation map on top of the thickness perturbation map (Fig.~\ref{fig:thick}), showing that there is a good match between the arm positions as calculated from the surface density and the thickness; the thickness of the HI layer is smaller in the arms than in the rest of the disk, as suggested previously \citep{HJK1982}.

%
%
%
%
%
%
%
%
%
%

%
%



\end{document}